\documentclass[runningheads]{llncs}
\usepackage{bm}
\usepackage[T1]{fontenc}
\usepackage{graphicx}
\usepackage{url} 
\usepackage{hyperref} 
\usepackage{cite}  

\usepackage{booktabs}
\usepackage{orcidlink}

\begin{document}
\title{How Instructional Sequence and Personalized Support Impact Diagnostic Strategy Learning}

\titlerunning{Impact of Instructional Sequence and Personalized Support}
%
\author{
Fatma Betül Güreş~\orcidlink{0000-0002-8664-1022}\inst{1} \and
Tanya Nazaretsky~\orcidlink{0000-0003-1343-0627}\inst{1} \and
Bahar Radmehr~\orcidlink{0009-0008-8034-4853}\inst{1} \and
\\ Martina Rau~\orcidlink{0000-0001-7204-3403}\inst{2} \and
Tanja Käser~\orcidlink{0000-0003-0672-0415}\inst{1}}

\authorrunning{F. B. Güreş et al.}
%

\institute{
EPFL, Switzerland \\
\email{\{fatma-betul.gures, tanya.nazaretsky, bahar.radmehr, tanja.kaeser\}@epfl.ch}
\and
ETH Zürich, Switzerland \\
\email{martina.rau@gess.ethz.ch}
}

\maketitle              
\begin{abstract}
Supporting students in developing effective diagnostic reasoning is a key challenge in various educational domains. Novices often struggle with cognitive biases such as premature closure and over-reliance on heuristics. Scenario-based learning (SBL) can address these challenges by offering realistic case experiences and iterative practice, but the optimal sequencing of instruction and problem-solving activities remains unclear.
This study examines how personalized support can be incorporated into different instructional sequences and whether providing explicit diagnostic strategy instruction before (I-PS) or after problem-solving (PS-I) improves learning and its transfer. We employ a between-groups design in an online SBL environment called PharmaSim, which simulates real-world client interactions for pharmacy technician apprentices. Results indicate that while both instruction types are beneficial, PS-I leads to significantly higher performance in transfer tasks. 


\end{abstract}

\section{Introduction}
\vspace{-5pt}

Higher-order thinking skills are essential for solving complex professional problems \cite{Shabrina2024,Abdelshiheed2024}. Diagnostic reasoning plays a critical role in healthcare, engineering, business, and education, helping practitioners draw evidence-based conclusions and plan interventions \cite{eriksen2024interprofessional,Graber2012,Zhang2022}. Yet, students often struggle with it, relying on heuristics and facing cognitive biases like premature closure \cite{kumar2011,Abdelshiheed2024}. 

Structured strategies like systematic data collection and thoughtful data interpretation \cite{elstein1978} can help reduce these issues~\cite{bowen2006,chi1981categorization,mcdaniel_campbell_hepworth_2005,norman2005,tversky1974}. Structured methods such as the LINDAFF checklist guide learners in covering key symptom dimensions, while strategies that consider interpersonal dynamics—like asking about family or caregivers—help uncover contextual factors relevant to diagnosis \cite{graber_checklists,pharmawiki_lindaaff,ncbi_family_dynamics}. Effective data interpretation requires generating a broad list of potential causes and evaluating their likelihood based on available evidence \cite{Graber2012,norman2005,croskerry2003}. This process is iterative: students generate hypotheses, identify missing information, and refine conclusions accordingly. Teaching such reasoning calls for structured support, realistic cases, and timely feedback \cite{eva2005,norman2005}.

Scenario-based learning (SBL) supports this need by offering realistic casework and immediate feedback \cite{Clark2012,LoiblEtAl2020}, though its open-ended nature can overwhelm learners without guidance \cite{Roll2018}. Personalized instruction helps—but how to best sequence instruction and problem-solving \cite{LoiblRollRummel2017} remains an open question.
Instruction-first (I-PS) prepares students with conceptual understanding before applying it \cite{LoiblRummel2014}, whereas problem-solving first (PS-I) encourages initial struggle, helping students recognize gaps in their understanding that instruction can then address \cite{Kapur2012,LoiblRollRummel2017,SinhaKapur2021}. This struggle has been shown to deepen learning and improve transfer \cite{SchwartzMartin2004}. While PS-I is effective in structured domains like math and physics \cite{Kapur2012,Saba2023}, its potential for teaching diagnostic strategies in SBL environments is underexplored \cite{LoiblRollRummel2017}. Moreover, little is known about how personalized feedback within these sequences affects learning transfer across contexts \cite{barnett2002when,salomon1989rocky}.
In this paper, we investigate how instructional sequencing—providing diagnostic strategy instruction before or after problem-solving—affects learning and transfer. Specifically, we ask: ``How does instructional sequencing affect students’ learning outcomes in acquiring diagnostic strategies across the transfer contexts?'' Using PharmaSim, an online SBL for pharmacy apprentices, we conduct a between-groups study with 80 participants. Results show that PS-I significantly improves far-transfer performance, highlighting the value of personalized feedback following problem-solving.
 
\vspace{-4mm}
\section{Methodology}
\begin{figure}[t]
    \includegraphics[width=1\textwidth]{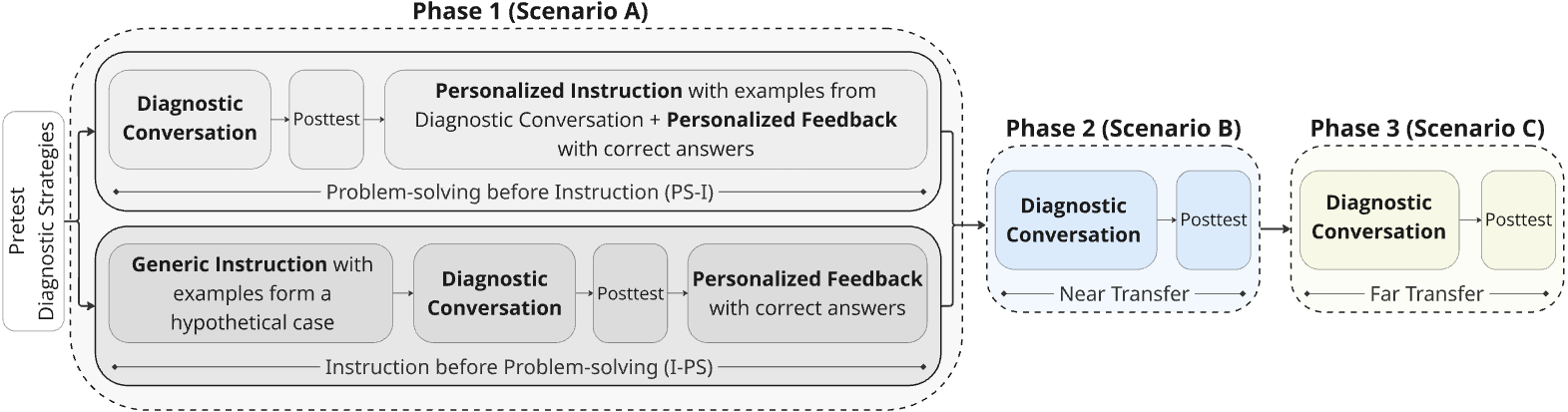}
    \caption{Experimental Design: students took pre-test on diagnostic strategies before engaging in a learning phase (Phase 1) (with instruction before or after a diagnostic activity) and two transfer phases (Phase 2 \& 3).}
    \label{fig:study_design}
    \vspace{-3mm}
\end{figure}

\vspace{-3mm}
Our between-group experimental study design is illustrated in Figure~\ref{fig:study_design}. Our study was divided into a pretest and three phases carefully designed to study the effect of instructional sequencing on students' ability to conduct learning diagnostic conversations and transfer their knowledge to different scenarios and contexts.. In all phases, students interacted with clients in PharmaSim, an SBL aimed at training pharmacy apprentices to master diagnostic reasoning skills. PharmaSim consists of two main modules and supports a range of client scenarios.
The \textit{Client Inquiry and Research Module} replicates client consultations (Fig.~\ref{fig:pharmasimResearch}, left). Students select individuals (e.g., main client or relatives) and inquiry topics (e.g., symptoms, age, allergies) from drop-down menus to gather data. They can also access a medicine compendium and relevant lecture notes.
After data collection, students enter the \textit{Diagnostic Decision Module}, where they list potential causes, justify them, and assign likelihood levels using labels like unlikely, likely, or very likely—mirroring structured diagnostic reasoning. We selected three scenarios (A, B, and C) focused on infant and maternal health to assess near and far transfer of diagnostic strategies.






\noindent \textbf{Scenario A} features a father seeking advice for his six-month-old baby (Client A) with 12-hour diarrhea. Students are expected to generate a list of possible causes, then rule out teething (due to age), dismiss viral infection (due to mild symptoms) and mother's medication (no interference with breastfeeding), and identify recent dietary changes as the likely cause.
\textbf{Scenario B} (near transfer) features the same possible causes as scenario A, but a different underlying issue (Client B). Here, more severe symptoms, no dietary changes, and maternal antibiotic use shift the likely cause to the mother’s condition. The scenario tests students’ ability to apply diagnostic strategies when cases look similar but differ in causality.
\textbf{Scenario C} (far transfer) introduces a mother (Client C1) with breastfeeding issues and concerns about her baby (Client C2). Both clients present distinct sets of symptoms, requiring students to identify different possible causes and determine interactions between them. This represents a far-transfer context with greater complexity and unfamiliarity.

\vspace{-4mm}
\subsection{Procedure}
\label{sec:ExperimentalDesign}
\vspace{-1mm}

Our study (see Fig.~\ref{fig:study_design}) started with a pretest on diagnostic strategies, followed by a learning session, in which both groups engaged in a diagnostic conversation either before or after receiving instructions. The subsequent phases then measured near-transfer and far-transfer again using simulated client scenarios.

\vspace{1pt} \noindent \textbf{Pretest.}
The pretest aimed to ensure that the knowledge levels of the experimental groups were similar. Participants were asked to provide a brief, 3–4 sentence response outlining how they would approach understanding a pharmacy customer’s problem and offering effective advice. Their responses were expected to reflect their prior knowledge about diagnostic strategies.

\vspace{1pt} \noindent \textbf{Phase 1.}
Both experimental groups engaged in diagnostic conversations, completed a post-test, and subsequently received personalized feedback on their performance. Additionally, both groups received the same instructional explanations regarding the rationale behind diagnostic strategies. However, the timing of instruction and the nature of the illustrative examples differed between the groups.

Participants in the PS-I group first interacted with Client A without prior instruction on diagnostic strategies. They then received instruction on diagnostic strategies, illustrated with examples drawn from their own prior interaction with Client A. In contrast, participants in the I-PS condition received the same instruction before interacting with Client A, using examples based on an hypothetical case, as no prior interaction had yet occurred. This structure enabled meaningful personalization within the constraints of each instructional sequence, with PS-I allowing instruction to incorporate students’ own problem-solving, and I-PS providing personalized feedback after the task.

\begin{figure}[t] 
\includegraphics[width=1\textwidth]{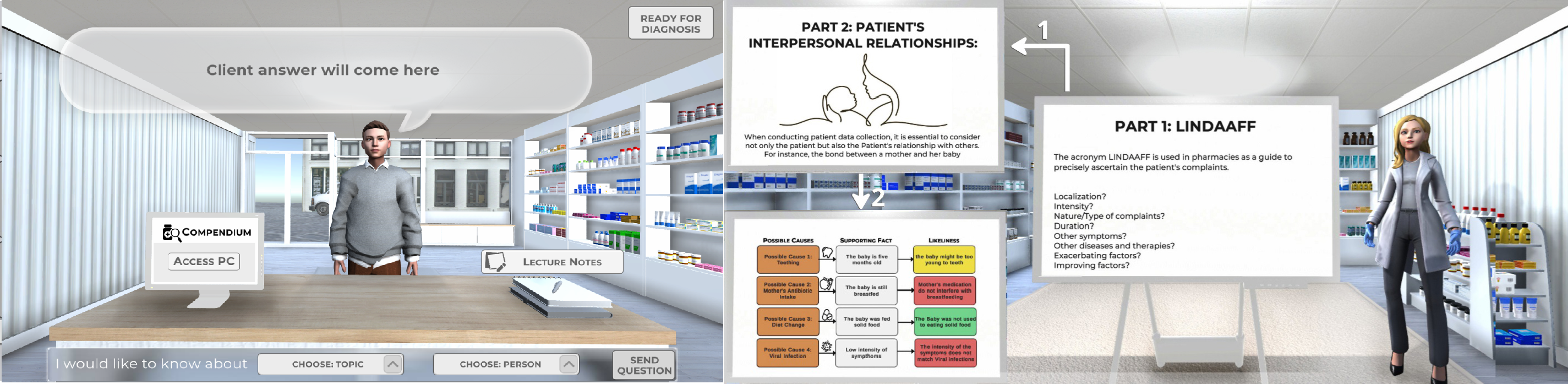} 
\caption{Client inquiry and research module (left), Feedback and Instruction (right).} \label{fig:pharmasimResearch} \vspace{-2mm} \end{figure}

For both groups, we integrated the instruction directly into PharmaSim as a virtual room, where students received the instruction and personalized feedback from a pharmacist character (see Fig.~\ref{fig:pharmasimResearch} (right)). By analyzing student interaction in the \textit{Client Inquiry and Research Module}, PharmaSim generates personalized \textit{performance-focused feedback} based on comparing student solutions to the correct diagnostic process. It highlights potential causes students have identified and points out critical causes they had overlooked. 


\vspace{1pt} \noindent \textbf{Phase 2 (Near Transfer).} The next phase aimed to assess whether and how students can adapt their diagnostic reasoning to slightly altered contexts (Client B) while retaining the core diagnostic strategies introduced in Phase 1. 
 
\vspace{1pt} \noindent \textbf{Phase 3 (Far Transfer).} Finally, participants tackled a more complex scenario involving two Clients, C1 and C2, which required them to integrate and apply their knowledge in a substantially different context. 

\vspace{-4mm}
\subsection{Measurement and Analysis}
\label{subsec:measurement-analysis}
\vspace{-2mm}
While participants engaged with PharmaSim, all interactions were logged, including the questions asked, the medicines and lecture notes accessed (\textit{Client Inquiry and Research Module}), and the diagnostic hypotheses generated and assessed (\textit{Diagnostic Decision Module}). The collected data included quantitative (click-stream) and qualitative data (pretest responses and final evaluative answers submitted during the client assessment).

We used diagnostic quality scores to assess students' acquisition and transfer of diagnostic strategies skills. The \textit{pre-test} assessing prior knowledge of diagnostic strategies was graded on a 0–4 scale based on four criteria: mentioning (i) the LINDAFF checklist, (ii) considering interpersonal relationships, (iii) generating a broad list of potential causes, and (iv) systematically prioritizing causes.   

We assessed student performance on the \textit{checklist}, \textit{interpersonal relationships}, and \textit{data interpretation} strategies. The \textit{checklist strategy} was assessed based on adherence to the LINDAFF framework, where each of the seven categories corresponded to a required diagnostic question. The LINDAFF score was calculated as the proportion of fulfilled categories. The \textit{interpersonal relationships strategy} was assessed on a scale of 0 to 3 by checking whether students asked three relevant questions about another person in the scenario (mother in scenarios A and B, baby in scenario C). The score was calculated as the proportion of fulfilled categories, quantifying consideration of interpersonal factors in diagnosis. Student performance in the \textit{data interpretation strategy} (\textit{Diagnostic Decision Module}) was assessed by evaluating the number of potential causes identified and explanation clarity. Scores included correctly identifying all possible causes (0 to 4 in scenarios A and B, 0 to 3 for C1 and C2 separately) and correctly assessing the likelihood and rationale (scored on the same scale). The total data interpretation strategy score was calculated as an average between C1 and C2 clients' scores. Below, this score is referred to as a post-test, as it reflects that students have internalized the diagnostic strategies and can effectively apply them to reach a coherent diagnostic conclusion. To compare scenarios, all scores were normalized as percentages of the total. These measures provide a multidimensional evaluation of the students learning outcomes.





\vspace{-5mm}
\subsection{Participants}
\vspace{-2mm}
Pharmacy technician apprentices (N = 80, 79 female) participated, reflecting the profession's gender distribution in Switzerland. All were second-year apprentices in a three-year vocational program with prior coursework in pharmaceutical sciences. Participation was voluntary and integrated into routine training, with informed consent obtained. Participants were randomly assigned to one of two instructional conditions: PS-I (n = 43) or I-PS (n = 37). All participants (in case of minors, their parents) gave informed consent and the study was approved by the university's ethics committee (Nr. 043-2021).

\vspace{-4mm}
\section{Results}
\vspace{-3mm}
We first verified that there were no difference in pretest scores between groups. We conducted a Mann–Whitney $U$ test (a Shapiro-Wilk test indicated that the pretest scores were not normally distributed (PS-I: $W = 0.796, p < .001$; I-PS: $W = 0.649, p < .001$)), confirming no significant differences in participants' prior knowledge of diagnostic strategies ($U = 843.5, p = .575$). Most students in both groups demonstrated familiarity with the LINDAFF strategy, while other methods were not explicitly mentioned or systematically applied in the pretest.

\begin{figure}[t!]
    \vspace{-5mm}
    \includegraphics[width=1\textwidth]{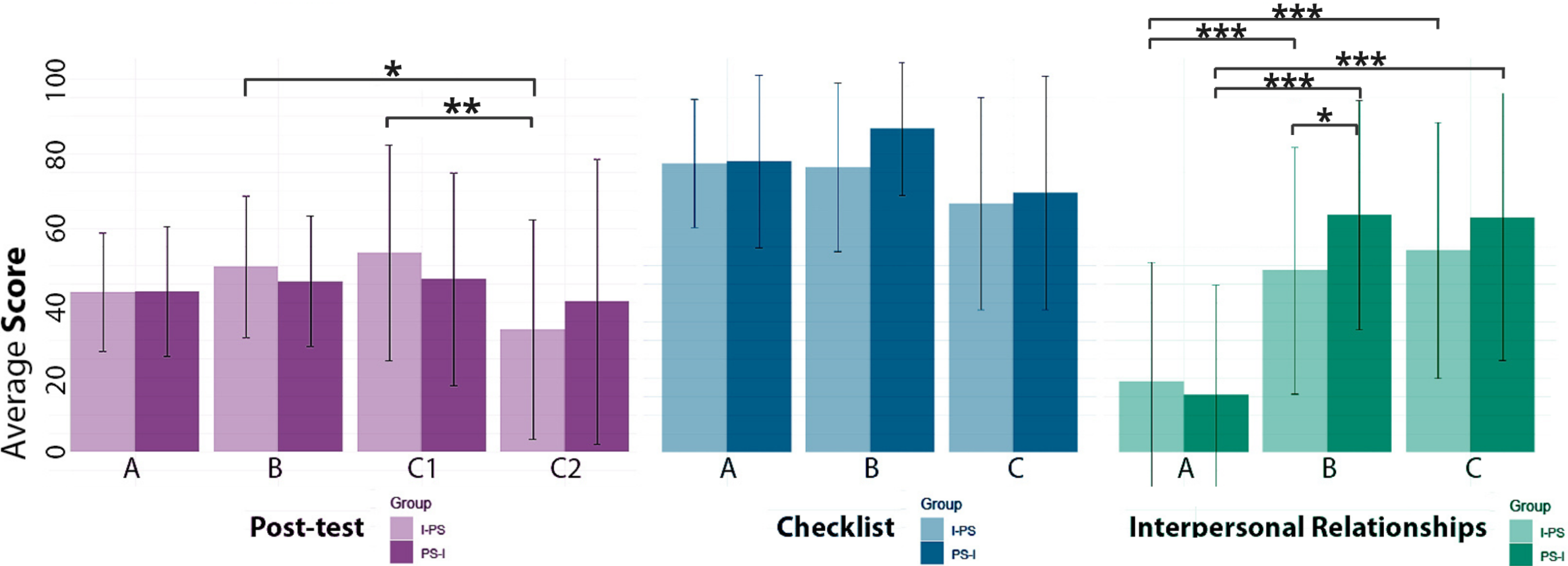}
    \vspace{-1em}
    \caption{Distribution of diagnostic strategy evaluation scores per scenario/client. Standard errors are heteroskedasticity robust (*** p < 0.001; ** p < 0.01; * p < .05).}
    \label{fig:RQ}
\end{figure}

We then used Mixed Linear Models (MLMs) to examine how instructional condition and scenario influenced strategy scores. Pretest scores were modeled as fixed effects; student ID was a random effect. We fit separate MLMs for each strategy. Checklist and interpersonal scores were scenario-based (A–C); post-test scores were client-based (A, B, C1, C2). We found no effect of the pretest, the experimental group condition, or the interactions between scenarios and experimental groups across all strategies. However, there was a significant effect of the diagnostic scenarios for checklist and interpersonal relationship strategies. The results of the post-hoc pairwise comparisons are presented in Fig.~\ref{fig:RQ}. There was a significant difference in the I-PS group post-test of client C2 ($\mu_{C2}^{I-PS}$ = 32.9, $\sigma_{C2}^{I-PS} = 29.5$) with both clients B ($\mu_{B}^{I-PS}$ = 49.7, $\sigma_{B}^{I-PS} = 18.9$), p = .0190 and C1 ($\mu_{C1}^{I-PS}$ = 53.4, $\sigma_{C1}^{I-PS} = 29.0$), p = .0023). These findings suggest that the I-PS group had difficulty applying or transferring what they learned to a more challenging scenario (Client C2), while PS-I maintained their performance.

Furthermore, students in both groups demonstrated a significant difference in scores for the interpersonal relationship strategy in scenarios A ($\mu_{A}^{I-PS}$ = 18.9, $\sigma_{A}^{I-PS} = 32.0$, $\mu_{A}^{PS-I}$ = 15.5, $\sigma_{A}^{PS-I} = 29.4$), B ($\mu_{B}^{I-PS}$ = 48.6, $\sigma_{B}^{I-PS} = 33.0$, $\mu_{B}^{PS-I}$ = 63.6, $\sigma_{B}^{PS-I} = 30.7$), and C ($\mu_{C}^{I-PS}$ = 54.1, $\sigma_{C}^{I-PS} = 34.3$, $\mu_{C}^{PS-I}$ = 62.8, $\sigma_{C}^{PS-I} = 38.2$), where all p < 0.001. 

The post-hoc pairwise comparisons showed no significant differences between conditions for all clients (p > 0.5) and scenarios besides a significant difference in the near transfer scenario (Client B) for interpersonal relationship strategy: I-PS group scores ($\mu_{B}^{I-PS}$ = 48.6, $\sigma_{B}^{I-PS} = 33.0$) were lower than PS-I group ($\mu_{B}^{PS-I}$ = 63.6, $\sigma_{B}^{PS-I} = 30.7$), p = .0438. This difference suggests that the personalized instruction in the PS-I condition was more effective in facilitating the acquisition of this strategy than the generic instruction provided in the I-PS condition. Students who acquired the strategy through connections to their prior experiences performed better in the near transfer scenario, as they were more adept at recognizing and applying the approach in a similar context.  

\vspace{-5mm}
\section{Discussion and Conclusions} 
\vspace{-3mm}

This study examined the impact of instructional sequencing and adaptive feedback on diagnostic strategy learning. Consistent with prior research \cite{LoiblRummel2014,LoiblEtAl2020,Sinha2021,DeCaro2022,SchwartzMartin2004}, PS-I proved more effective than I-PS, particularly for far-transfer tasks. PS-I students maintained stable diagnostic reasoning under complex conditions, while I-PS students showed performance declines. Engaging in problem-solving before instruction helped PS-I students build an initial understanding, later refined through instruction and personalized feedback, aligning with \cite{Saba2023} and extending it through analysis of learning patterns and information-seeking behaviors.
Our study also showed that the I-PS group did not outperform a "no-instruction" baseline, contrasting with \cite{Saba2023}. Performance gains for I-PS students only appeared after Phase 2, suggesting that feedback, not instruction alone, drove meaningful learning—likely due to the cognitive demands of diagnostic tasks. Although our sample size was sufficient, the pharmacy context may limit generalizability. Future research should replicate these methods across varied domains and integrate qualitative analyses to better understand students' reasoning. 

\vspace{1pt} \noindent  \textbf{Acknowledgements.}
This study was funded by the Swiss State Secretariat for Education, Research and Innovation SERI and the Jacobs Foundation.

\bibliographystyle{splncs04}
\bibliography{references}

\end{document}